\documentstyle[twoside,fleqn,espcrc2]{article}

\newcommand{\AmS}{{\protect\the\textfont2
  A\kern-.1667em\lower.5ex\hbox{M}\kern-.125emS}}
\title{Spectator Effects in
      Inclusive Decays of Beauty Hadrons
\thanks{talk presented by Massimo Di Pierro}}

\author{{ \bf UKQCD Collaboration}: Massimo Di Pierro 
	\address{Department of Physics and Astronomy,
        University of Southampton, SO17 1BJ, United Kingdom }
        and Chris T. Sachrajda$^{\rm a}$}

\begin{document}
\begin{abstract}
We evaluate the matrix elements of the four-quark operators which
contribute to the lifetimes of $B$-mesons and the $\Lambda_b$-baryon. We
find that the spectator effects are not responsible for the discrepancy
between the theoretical prediction and experimental measurement of the
ratio of lifetimes $\tau(\Lambda_b)/\tau(B)$.
\end{abstract}

\maketitle

\section{INTRODUCTION}

Inclusive decays of heavy hadrons can be studied in the framework of the
heavy quark expansion, in which, for example, lifetimes are computed as
series in inverse powers of the mass of the $b$-quark~\cite{Shif}. For
an arbitrary hadron $H$
\begin{equation}
\tau ^{-1}(H)=\frac{G_F^2m_b^5}{192\pi ^3}\frac{|V_{cb}|^2}{2m_H} 
\sum_{i\geq 0} c_i m^{-i}_b
\label{ope}
\end{equation}
where
\begin{itemize}
\item  $c_0$ corresponds to the decay of a free-quark and is
universal.
\item  $c_1$ is zero because the operators of dimension
four can be eliminated using the equations of motion.
\item  $c_2$ can be estimated and is found to be small.
\item  $c_3$ contains a contribution proportional to 
\begin{equation}
\left\langle H\right| \ \overline{b}\Gamma q\ \overline{q}\widetilde{\Gamma }%
b\ \left| H\right\rangle   \label{4quark}
\end{equation}
and is therefore the first term in the expansion to which the
interaction between the heavy and the light quark(s) contribute. 
Although this is an $O(m_b^{-3})$ correction, it may be significant
since it contains a phase-space enhancement.
\end{itemize}

The aim of our lattice simulation is to compute 
$c_3$ for $B$-mesons and the $\Lambda_b$-baryon, in order to check whether
spectator effects contribute significantly to the ratios of lifetimes for
which the experimental values are:
\begin{eqnarray}
\frac{\tau (B^{-})}{\tau (B^0)}&=&1.06\pm 0.04 \label{experiment1} \\
\frac{\tau (\Lambda _b)}{\tau (B^0)}&=&0.78\pm 0.04\ .\label{experiment2}
\end{eqnarray}

The discrepancy between the experimental value in 
eq.~(\ref{experiment2}) and the theoretical prediction of
$\tau(\Lambda_b)/\tau(B^0)=0.98$ (based on the Operator Product
Expansion in eq.~(\ref{ope}) including terms in the sum up to those of
$O(m_b^{-2})$) is a major puzzle. It is therefore particularly important
to compute the $O(m_b^{-3})$ spectator contributions to this ratio.

The ratios in eqs.~(\ref{experiment1}) and (\ref{experiment2}) can be 
expressed in terms of 6 matrix elements:
\begin{eqnarray}
\frac{\tau (B^{-})}{\tau (B^0)} &=&a_0+a_1\varepsilon _1
+a_2 \varepsilon _2 +a_3 B_1+a_3 B_2\\
\frac{\tau (\Lambda _b)}{\tau (B^0)} &=&b_0 +b_1 \varepsilon _1
+ b_2 \varepsilon _2 + b_3 L _1 +b_4 L _2
\end{eqnarray}
where 
\footnote{
In terms of the parameters $\widetilde B$ and $r$ introduced in
ref.~\cite{CTS} 
\begin{eqnarray*}
r &=& -6 L_1 \\
\widetilde B &=& -2 L_2 / L_1 - 1/3
\end{eqnarray*}}
\begin{eqnarray}
B_1 &\equiv & \frac 8{f_B^2m_B}\frac{\left\langle B\right| \ 
\overline{b}\gamma ^\mu Lq\ \overline{q}\gamma _\mu Lb\ \left|
B\right\rangle }{2m_B}  \label{B1} \\
B_2 &\equiv & \frac 8{f_B^2m_B}\frac{\left\langle B\right| \ 
\overline{b}Lq\overline{\ q}Rb\ \left| B\right\rangle }{2m_B} 
\label{B2}\\
\varepsilon _1 &\equiv & \frac 8{f_B^2m_B}\frac{\left\langle
B\right| \ \overline{b}\gamma ^\mu Lt^aq\ \overline{q}\gamma _\mu Lt^ab\
\left| B\right\rangle }{2m_B} \label{epsilon1}\\
\varepsilon _2 &\equiv & \frac 8{f_B^2m_B}\frac{\left\langle
B\right| \ \overline{b}Lt^aq\ \overline{q}Rt^ab\ \left| B\right\rangle }{2m_B%
} \label{epsilon2}\\
L _1 &\equiv&\frac 8{f_B^2m_B}\frac{\left\langle \Lambda
\right| \ \overline{b}\gamma ^\mu Lq\ \overline{q}\gamma _\mu Lb\ \left|
\Lambda \right\rangle }{2m_\Lambda } \label{lambda1} \\
L _2 &\equiv & \frac 8{f_B^2m_B}\frac{\left\langle \Lambda
\right| \ \overline{b}\gamma ^\mu Lt^aq\ \overline{q}\gamma _\mu Lt^ab\
\left| \Lambda \right\rangle }{2m_\Lambda }  \label{lambda2}
\end{eqnarray}
and the coefficients $a_i$ and $b_i$ are given by
\begin{center}
\begin{tabular}{llll}
\hline
& value & & value \\ \hline
$a_0$ & $+1.00$  & $b_0$ & $+0.98$ \\
$a_1$ & $-0.697$ & $b_1$ & $-0.173$ \\
$a_2$ & $+0.195$ & $b_2$ & $+0.195$ \\
$a_3$ & $+0.020$ & $b_3$ & $+0.030$ \\
$a_4$ & $+0.004$ & $b_4$ & $-0.252$ \\
\hline
\end{tabular}
\end{center}

The values in the table correspond to operators 
renormalized in a continuum renormalization scheme at the scale $\mu=m_B$.
Their matrix elements are obtained from those in the lattice 
regularization by perturbative matching (see the appendix).

\section{$B$ DECAY}

The matrix elements $B_1,B_2,\varepsilon _1,\varepsilon _2$ are computed
on a $24^3\times 48$ lattice at $\beta =6.2$ (corresponding to a lattice
spacing $a^{-1}=2.9(1)$~GeV) using the tree-level improved SW action 
for three values of $\kappa =0.14144$, $0.14226$, $0.14262$ and are then
extrapolated to the chiral limit ($\kappa_c=0.14315$)~\cite{me1}. We
find
\begin{eqnarray}
B_1 &=&1.06\pm 0.08 \label{vB1} \\
B_2 &=&1.01\pm 0.06 \label{vB2} \\
\varepsilon _1&=&-0.01\pm 0.03 \label{vepsilon1} \\  
\varepsilon _2&=&-0.02\pm 0.02 \label{vepsilon2}
\end{eqnarray}
which implies that
\begin{equation}
\frac{\tau (B^{-})}{\tau (B^0)}=1.03\pm 0.02\pm 0.03
\end{equation}
in agreement with the experimental value (\ref{experiment1}).

\section{$\Lambda $ DECAY}

The computation the baryonic matrix elements $L_1$ and $L_2$ 
is a little more difficult. We have performed an exploratory study in 
which the light quark propagators are computed using a stochastic method
\cite{CM} based on the relation
\begin{equation}
M_{ij}^{-1}=\int [d\phi ](M_{jk}\phi _k)^{*}\phi _i e^{-\phi
_i^{*}(M^{+}M)_{ij}\phi _j}
\end{equation}
(rather then using ``extended'' propagators).

The matrix elements are computed on a $12^3\times 24$ lattice 
at $\beta =5.7$ (corresponding to a lattice spacing $%
a^{-1}=1.10(1)$ GeV) for two values of $\kappa$. We therefore do
not attempt an extrapolation to the chiral limit ($\kappa_c=0.14351$)
but present results seperately for each value of $\kappa$.
We find:
\begin{equation}
L _1=\cases{
-0.30\pm 0.03 & ($\kappa=0.13843$) \cr
-0.22\pm 0.03 & ($\kappa=0.14077$)
} \label{vlambda1} 
\end{equation}
\begin{equation}
L _2=\cases{
0.23\pm 0.02 & ($\kappa=0.13843$) \cr
0.17\pm 0.02 & ($\kappa=0.14077$) \ ,
}\label{vlambda2}
\end{equation}
which implies that (neglecting the systematic error due to the
chiral extrapolation)
\begin{equation}
\frac{\tau (\Lambda _b)}{\tau (B^0)}=\cases{
0.91\pm 0.01 & ($\kappa=0.13843$) \cr
0.93\pm 0.01 & ($\kappa=0.14077$) \ .
}
\end{equation}
These results imply that spectator effects are not sufficiently large to
explain the discrepancy between the theoretical prediction and
experimental result in eq.~(\ref{experiment2}).

\begin{table*}[hbt]
\setlength{\tabcolsep}{0.5pc}
\newlength{\digitwidth} \settowidth{\digitwidth}{\rm 0}
\catcode`?=\active \def?{\kern\digitwidth}
\caption{Lattice perturbative coefficients and operators}
\label{table1}
\begin{tabular*}{\textwidth}{@{}llll}
\hline
$p_i$ & $P_i$ & $q_i$ & $Q_i$ \\
\hline 
$-\frac43\log(\lambda^2 a^2)+8.46$ & 
$\overline{b}\Gamma q\overline{q}\widetilde{\Gamma }b$ & $+1
$ & $\overline{b}\Gamma q\overline{b}\widetilde{\Gamma }q$ \\ 
$-\log(\lambda^2 a^2)+6.19$ & $\overline{b}t^a\Gamma t^aq
\overline{q}\widetilde{\Gamma }b+
\overline{b}\Gamma q\overline{q}t^a\widetilde{\Gamma }t^ab$ & $+1$ & $
\overline{b}t^a\Gamma t^aq\overline{b}\widetilde{\Gamma }q+\overline{b}
\Gamma q\overline{b}t^a\widetilde{\Gamma }t^aq$ \\ 
$-6.89$ & $\overline{b}t^a\gamma ^0\Gamma \gamma ^0t^aq\overline{q}
\widetilde{\Gamma }b+\overline{b}\Gamma q\overline{q}t^a\gamma ^0\widetilde{
\Gamma }\gamma ^0t^ab$ & $+1$ & $\overline{b}t^a\gamma ^0\Gamma \gamma ^0t^aq
\overline{b}\widetilde{\Gamma }q+\overline{b}\Gamma q\overline{b}t^a\gamma ^0
\widetilde{\Gamma }\gamma ^0t^aq$ \\ 
$2\log(\lambda^2 a^2)-4.53$ & $\overline{b}t^a\Gamma q\overline{q}\widetilde{\Gamma }t^ab$
& $-1$ & $\overline{b}t^a\Gamma q\overline{b}t^a\widetilde{\Gamma }q$ \\ 
$\log(\lambda^2 a^2)-6.19$ & $\overline{b}\Gamma t^aq
\overline{q}\widetilde{\Gamma }t^ab+
\overline{b}t^a\Gamma q\overline{q}t^a\widetilde{\Gamma }b$ & $-1$ & $
\overline{b}\Gamma t^aq\overline{b}t^a\widetilde{\Gamma }q+\overline{b}
t^a\Gamma q\overline{b}\widetilde{\Gamma }t^aq$ \\ 
$-6.89$ & $\overline{b}\Gamma \gamma ^0t^aq\overline{q}\widetilde{
\Gamma }\gamma ^0t^ab+\overline{b}t^a\gamma ^0\Gamma q\overline{q}t^a\gamma
^0\widetilde{\Gamma }b$ & $+1$ & $\overline{b}\Gamma \gamma ^0t^aq\overline{b
}t^a\gamma ^0\widetilde{\Gamma }q+\overline{b}t^a\gamma ^0\Gamma q\overline{b
}\widetilde{\Gamma }\gamma ^0t^aq$ \\ 
$-\log(\lambda^2 a^2)+5.12$ & $\overline{b}\Gamma 
t^aq\overline{q}\gamma ^0\widetilde{
\Gamma }b$ & $-1$ & $\overline{b}\Gamma t^aq\overline{b}\widetilde{\Gamma }
t^aq$ \\ 
$-\frac14 \log(\lambda^2 a^2)+1.05$ & $\overline{b}\Gamma \sigma ^{\mu \nu }t^aq\overline{q}
t^a\sigma ^{\nu \mu }\widetilde{\Gamma }b$ & $-1$ & $\overline{b}\Gamma
\sigma ^{\mu \nu }t^aq\overline{b}\widetilde{\Gamma }\sigma ^{\mu \nu }t^aq$
\\ 
$-2.43$ & $\overline{b}\Gamma \gamma ^\mu t^aq\overline{q}t^a\gamma
^\mu \widetilde{\Gamma }b$ & $+1$ & $\overline{b}\Gamma \gamma ^\mu t^aq
\overline{b}\widetilde{\Gamma }\gamma ^\mu t^aq$ \\
\hline
\multicolumn{4}{@{}p{120mm}}{The analytic expression for $p_i$ can 
be found in \cite{me1}.}
\end{tabular*}
\end{table*}

\section{CONCLUDING REMARKS}

We find that the matrix elements of the 4-quark operators
(\ref{vB1}-\ref{vepsilon2}) satisfy the vacuum saturation hypothesis
remarkably well. A similar feature is true for the $\Delta B=2$
operators  which contribute to $B$--$\bar B$ mixing. We do not have a
good understanding yet of this phenomenon. 

The results  for the mesonic matrix elements lead to a prediction for 
the ratio of the lifetimes of the charged and neutral mesons which  is
in agreement with the experimental result in eq.~(\ref{experiment1}).
Our study of the baryonic matrix elements indicates that  spectator
effects are not sufficiently large to explain the experimental ratio of
lifetimes in eq.~(\ref{experiment2}). This discrepancy between  the
theoretical prediction and the experimental measurement remains an
important problem to solve. We do stress, however,  that our
calculations are exploratory, and a more precise simulation  is
necessary, in particular to allow for a reliable extrapolation  to the
chiral limit.

\section*{ACKNOWLEDGEMENTS}
It is a pleasure to thank Chris Michael, Hartmut Wittig, Jonathan Flynn,
Luigi Del Debbio, Giulia De Divitiis, Vicente Gimenez and Carlotta
Pittori  for many helpful discussions. This work was supported by PPARC
grants GR/L29927,  GR/L56329 and GR/I55066.

\section*{APPENDIX: 1-loop lattice perturbative corrections to 
4-quark operators}

The most difficult component of the evaluation of the 1-loop
perturbative matching between four-quark lattice operators and those  
renormalized in a continuum scheme is the perturbative expansion on
lattice. In this appendix we present the corresponding results for
generic operators $P_0$ and $Q_0$ (whose matrix elements contribute to
lifetimes and $B$--$\bar B$ mixing respectively):
\begin{eqnarray}
P_0&\equiv&\overline{b}\Gamma q\,\,\overline{q}\widetilde{\Gamma }b,
\qquad (\Delta B=0) \label{db0} \\
Q_0&\equiv&\overline{b}\Gamma q\,\,\overline{b}\widetilde{\Gamma }q,
\qquad (\Delta B=2)\ . \label{db2}
\end{eqnarray}
$\Gamma \otimes \widetilde{\Gamma}$ represents an arbitrary
spinor and color tensor.
These operators mix under renormalization with other 4-quark operators,
listed in table~\ref{table1}~\footnote{These results were also obtained independently
in ref.~\cite{gimenez}.}:
\begin{eqnarray}
P_0^{\mathrm{1\,loop}} &=&P_0+\frac{\alpha _s(a^{-1})}{4\pi }\sum_i p_i P_i \\
Q_0^{\mathrm{1\,loop}} &=&Q_0+\frac{\alpha _s(a^{-1})}{4\pi }\sum_i p_i q_i Q_i
\ .\end{eqnarray}
The Feynman rules correspond to the tree-level SW-improved action for
massless light quarks and with a small gluon-mass ($\lambda$)  as the
infrared regulator. The dependence on $\lambda$, of course, cancels when
the corresponding continuum calculation is combined with the lattice
one.

\end{document}